\documentclass[12pt,a4paper]{article}

\usepackage{amssymb}
\usepackage{setspace}
\usepackage{indentfirst}

\usepackage{graphicx}

\usepackage{url}

\begin{document}

\begin{titlepage}
\begin{center}

\vspace*{25mm}

\begin{spacing}{1.7}
{\LARGE\bf Late-time data require smaller sound horizon at recombination}
\end{spacing}

\vspace*{25mm}

{\large
Noriaki Kitazawa
}
\vspace{10mm}

Department of Physics, Tokyo Metropolitan University,\\
Hachioji, Tokyo 192-0397, Japan\\
e-mail: noriaki.kitazawa@tmu.ac.jp

\vspace*{25mm}

\begin{abstract}
The Hubble constant problem is that
 the values of Hubble constant from the observation of
 cosmic microwave background assuming the $\Lambda$CDM model
 disagrees with the values from direct measurements.
This problem suggests some new physics beyond the $\Lambda$CDM model.
Typically there are two ways of reconciliation:
 one is the realization of smaller value of sound horizon at recombination,
 and the other is the modification of the way of expansion of the universe
 after recombination.
In this letter we examine the latter possibility
 by comparing two typical phenomenological dark energy models
 with the distance-redshift relation provided by
 Pantheon catalogue of supernova observations and
 galaxy surveys by BOSS and eBOSS collaborations. 
Though these phenomenological dark energy models
 globally fit observations better than the $\Lambda$CDM model,
 they are strongly disfavored by the distance-redshift relation
 as almost the same level as the $\Lambda$CDM model
 defined by cosmic microwave background observations.
The distance-redshift relation strongly suggests
 some new physics which realize smaller value of sound horizon at recombination.
\end{abstract}

\end{center}
\end{titlepage}

\doublespacing

\section{Introduction}
\label{sec:introduction}

The Hubble constant problem can be described as that
 the values of Hubble constant from the observation of
 cosmic microwave background (CMB) assuming the $\Lambda$CDM model
 contradict with the values from direct measurements
 (see \cite{Knox:2019rjx,DiValentino:2021izs,Schoneberg:2021qvd}
  for review and references therein).
The typical value of Hubble parameter
 from CMB observation assuming the $\Lambda$CDM model
 is $H_0 = 67.4 \pm 0.5 \,\, \rm{Km/s/Mpc}$
 by the Planck collaboration \cite{Planck:2018vyg}
 ({\it i.e.} the value with inverse distance ladder),
 and the typical value by direct measurements
 is $H_0 = 73.2 \pm 1.3 \,\, \rm{Km/s/Mpc}$ \cite{Riess:2020fzl}
 ({\it i.e.} the value with local distance ladder).
For a review of other direct measurements see \cite{Freedman:2023jcz}, for example.
The difference of these two values is more than $4$ standard deviations,
 which is the level that we can not neglect.
The Pantheon catalogue
 of the measurements of type Ia supernovae \cite{Pan-STARRS1:2017jku}
 acts very important rolls for many precise direct measurements.
The observations of baryon acoustic oscillation (BAO)
 in galaxy surveys by BOSS and eBOSS collaborations \cite{eBOSS:2020yzd}
 also acts important rolls
 to understand the origin of the difference of the values of Hubble constant.
In this letter
 we take the present results of these experiments at face value,
 though the experimental results could change
 as shown in History Plot in \cite{ParticleDataGroup:2020ssz}.
What the theorist should do
 may not make effort to reconcile the above two very different values exactly,
 but may find some natural physics which can at least naturally explain
 the difference of the results with local distance ladder and inverse distance ladder.

We describe
 a simple illustrative argument
 how we obtain Hubble constant from CMB observations,
 which is given in \cite{Knox:2019rjx}, in a more practical way.
The angular scale of BAO in CMB, $\theta_s^*$, is precisely measured,
 where the subscript means sound horizon and asterisk means that
 the value is at the time of ``last scattering''
 when the value of optical depth of baryon-photon plasma becomes one.
This quantity is theoretically described as
\begin{equation}
 \theta_s^* = r_s^* / d_A^*,
\label{AngularSoundHorizon}
\end{equation}
 where $r_s^*$ is the scale of sound horizon at the last scattering
 and $d_A^*$ is the angular diameter distance to the last scattering surface
 in comoving coordinate
\begin{equation}
 d_A^* = \int_{t^*}^{t_0} \frac{1}{a(t)} dt
       = \int_0^{z^*} \frac{dz}{H(z)},
\label{DistanceToLLS}
\end{equation}
 where $z^*$ is the redshift at the time of last scattering $t^*$.
In $\Lambda$CDM model
 the Hubble parameter as a function of redshift $z$ is described as
\begin{eqnarray}
 H(z) &=& H_0 \sqrt{\Omega_\Lambda + \Omega_m (1+z)^3 + \Omega_\gamma (1+z)^4}
\nonumber\\
 &=& (H_0/h) \sqrt{\omega_\Lambda + \omega_m (1+z)^3 + \omega_\gamma (1+z)^4}
\label{HubbleParameter}
\end{eqnarray}
 with the sum of the present energy densities of vacuum, matter and photons
 satisfying $\Omega_\Lambda + \Omega_m + \Omega_\gamma = 1$
 assuming the flat universe ($K=0$),
 where the dimensionless physical energy density
 is defined as $\omega \equiv \Omega h^2$ with $H_0 = 100 \, h \,\, \rm{Km/s/Mpc}$
 and we have neglected neutrino contribution for simplicity.
The scale of sound horizon at the last scattering is described as
\begin{equation}
 r_s^* = \int_{z^*}^\infty \frac{dz}{H(z)} c_s(z),
\label{SoundHorizon}
\end{equation}
 where
\begin{equation}
 c_s(z) = \frac{1}{\sqrt{3}}
          \left(1+\frac{1}{1+z}\frac{3\omega_b}{4\omega_\gamma}\right)^{-1/2}
\end{equation}
 is the sound velocity in baryon-photon plasma, which can be obtained
 from the system of equations of perturbations
 in case that the wavelength of the perturbations is well within the horizon
 and assuming small effects of the expansion of the universe.
Here, $\omega_b$ and $\omega_\gamma$
 are present physical energy densities of baryons and photons.

Since $\omega_b$ and $\omega_m$ are obtained by CMB observations
 and $\omega_\gamma$ is obtained from present CMB temperature
 assuming blackbody radiation,
 we can evaluate the integral of eq.(\ref{SoundHorizon}).
Because the value of $z^*$ is of the order of $10^3$,
 we can safely ignore $\omega_\Lambda$ in eq.(\ref{HubbleParameter})
 in this evaluation.
Note that the resultant value is independent from the value of $H_0$ or $h$.
The value of the sound horizon $r_s^* = 147.09 \pm 0.26$ [Mpc]
 is given by the Planck collaboration
 with the other method \cite{Planck:2018vyg}.
Once the value of $r_s^*$ is obtained
 we have a constraint for $d_A^*$ through eq.(\ref{AngularSoundHorizon}).
Then the integration of eq.(\ref{DistanceToLLS})
 determine the value of $h$ with the condition of
 $\omega_\Lambda = h^2 - \omega_m - \omega_\gamma$.
The smaller values of $r_s^*$
 result larger $h$ for smaller $d_A^*$ and vice vasa
 to keep the observed value of $\theta_s^*$ constant.
Note that in the integral of eq.(\ref{DistanceToLLS})
 the whole history of the expansion of the universe
 from the time of last scattering to the present contributes.

There are many models
 to realize smaller values of $r_s^*$ and larger value of $h$
 by introducing new physics beyond the $\Lambda$CDM model
 at the early time between the time of recombination
 and around the time of matter-radiation equality
 (see \cite{Poulin:2018cxd,Niedermann:2020dwg,Escudero:2019gvw,Brinckmann:2022ajr}
 for some typical proposals and \cite{DiValentino:2021izs,Schoneberg:2021qvd}
 for more complete lists of many efforts).
In this letter we examine the possibility to realize larger value of $h$
 through the modification of the way of expansion of the universe after recombination,
 which affects $d_A^*$, without changing $r_s^*$.
Typically, changing the way of expansion of the universe
 can be realized by phenomenologically introducing
 the time dependence of the dark energy.
Note that in the $\Lambda$CDM model the dark energy,
 or the cosmological constant, has been phenomenologically introduced
 to only realize recent accelerated expansion of the universe.
The Hubble constant problem
 may be giving some information on the dark sector
 beyond that in the $\Lambda$CDM model.

In the next section
 we introduce the experimental result of distance-redshift relation
 from Pantheon catalogue of type Ia supernovae
 and BAO data by the BOSS and eBOSS collaborations.
The distance-redshift relation
 is compared with the prediction of $\Lambda$CDM model
 whose parameters are fixed by the Planck CMB observations,
 and we see how strongly the model is disfavored.
Though this is a naive method
 only using the distance-redshift relation,
 it gives information which is not apparent
 in a global fit with all the available data.
We can not accept the model,
 if it strongly contradict with a part of data,
 even if the other data can be reasonably reproduced.
In section \ref{sec:PhDE}
 we examine two phenomenological dark energy models:
 ``emergent dark energy model`` \cite{Li:2020ybr,Yang:2021eud}
 and ``transitional dark energy model'' \cite{Zhou:2021xov}.
We see that the present distance-redshift relation does not support these models,
 though the statistical significances of the global fits of these models
 with all the available data are better than that of $\Lambda$CDM model.
In the last section
 we give discussions and conclusions
 that the present data strongly suggest, or even require,
 some new physics which realizes smaller value of sound horizon at recombination.

\section{The $\Lambda$CDM model and distance-redshift relation}
\label{sec:LCDM}

The Pantheon catalogue of type Ia supernovae \cite{Pan-STARRS1:2017jku}
 gives average apparent magnitudes of type Ia supernovae in $40$ redshift bins,
 which is open for public.\footnote{The data are in public at
 \url{https://archive.stsci.edu/prepds/ps1cosmo/index.html}.}
A simple formula of
 the apparent magnitude of a type Ia supernova at redshift $z$ is
\begin{equation}
 m(z) = 5 \log_{10} \left( d_L(z) \, \mbox{[Mpc]} \right) + 25 + M,
\end{equation}
 where $d_L(z)$
 is the luminosity distance to the supernova $d_L(z) = (1+z) r(z)$ with
\begin{equation}
 r(z) \equiv \int_0^z \frac{dz'}{H(z')}
\label{Distance}
\end{equation}
 which is the light propagation distance
 and $M$ is the common absolute magnitude of type Ia supernovae.
We take the value of $M = -19.263$ [mag]
 which has been obtained in \cite{Kitazawa:2021ycu}
 by the fit of the same Pantheon data.
Then we have distance-redshift relation
\begin{equation}
 r(z) = \frac{1}{1+z} \, 10^{(m(z)-(25+M))/5} \,\, \mbox[\rm Mpc].
\end{equation}

The observations of BAO by BOSS and eBOSS \cite{eBOSS:2020yzd}
 gives the ratios of $D_M(z)/r_d$ at six values of effective redshifts,
 which is open for public,\footnote{The data are in public at
 \url{https://www.sdss4.org/science/final-bao-and-rsd-measurements/}.}
 where $D_M(z) = r(z)$ is the angular diameter distance in comoving frame
 and
 we replace the original $r_d$,
 the value of sound horizon at the end of baryon drag epoch,
 by $r_s^*$ in the following for simplicity.
The $2\%$ difference of these values \cite{Knox:2019rjx}
 is negligibly small in our arguments.
 
\begin{figure}[t]
\centering
\includegraphics[width=50mm]{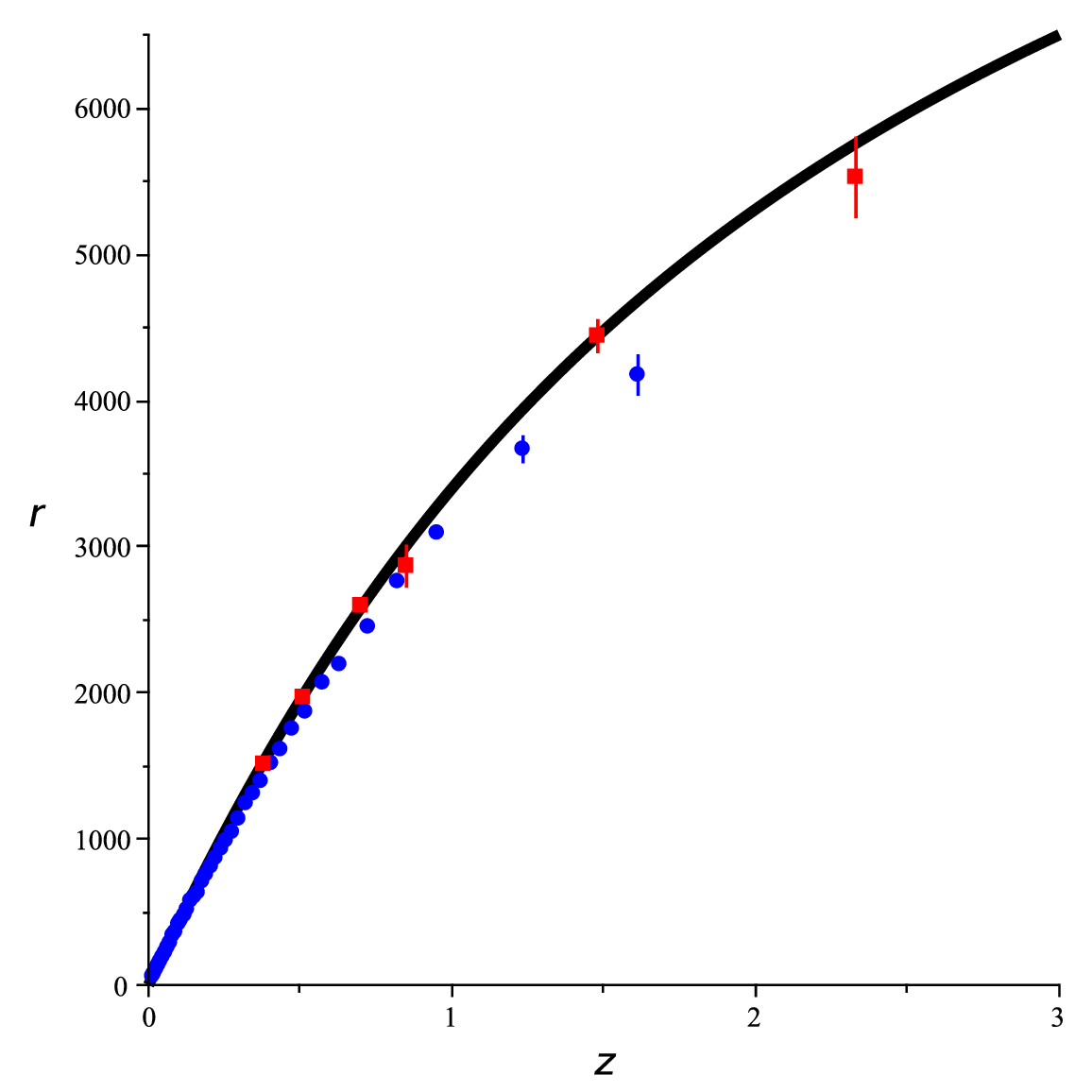}
\quad
\includegraphics[width=50mm]{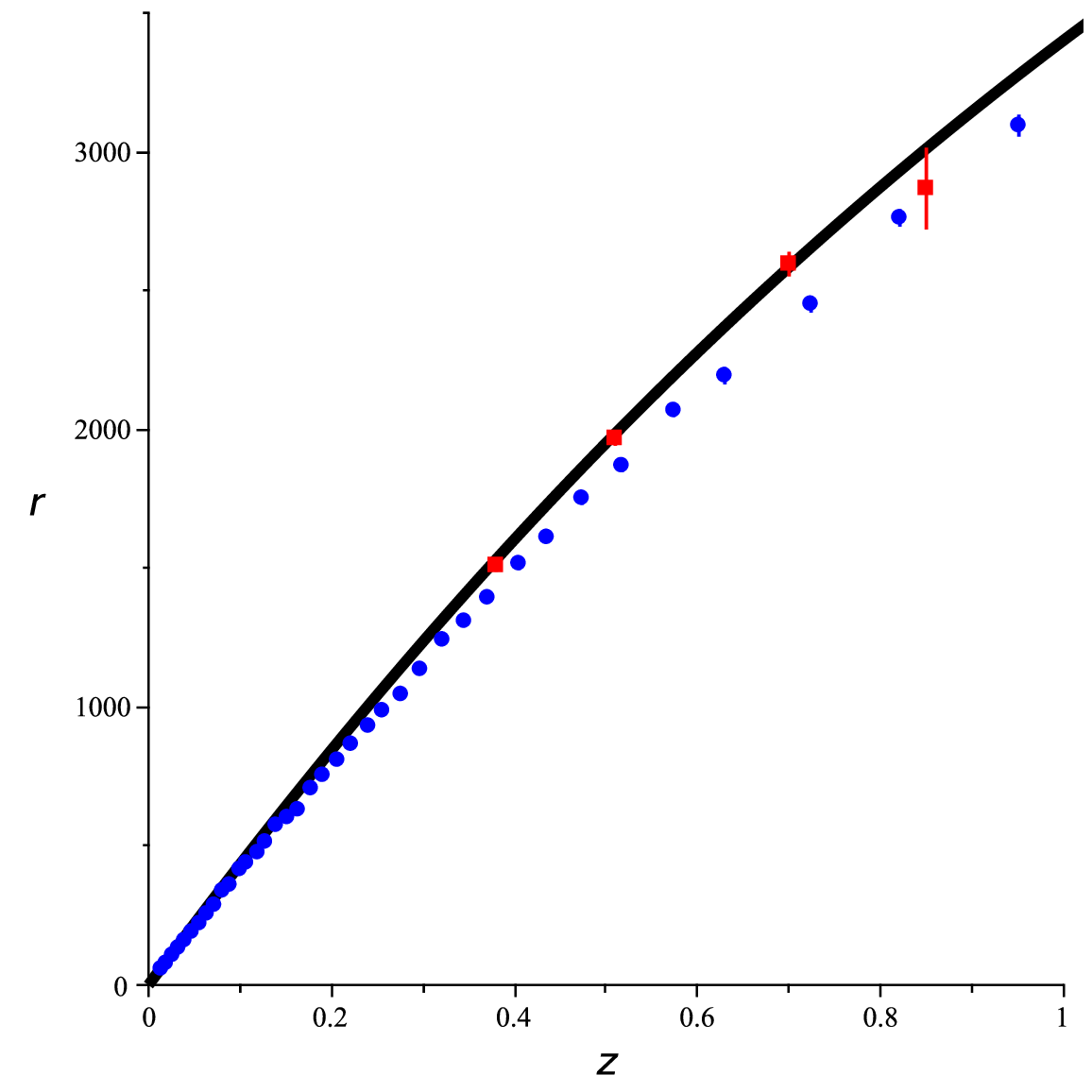}
\quad
\includegraphics[width=50mm]{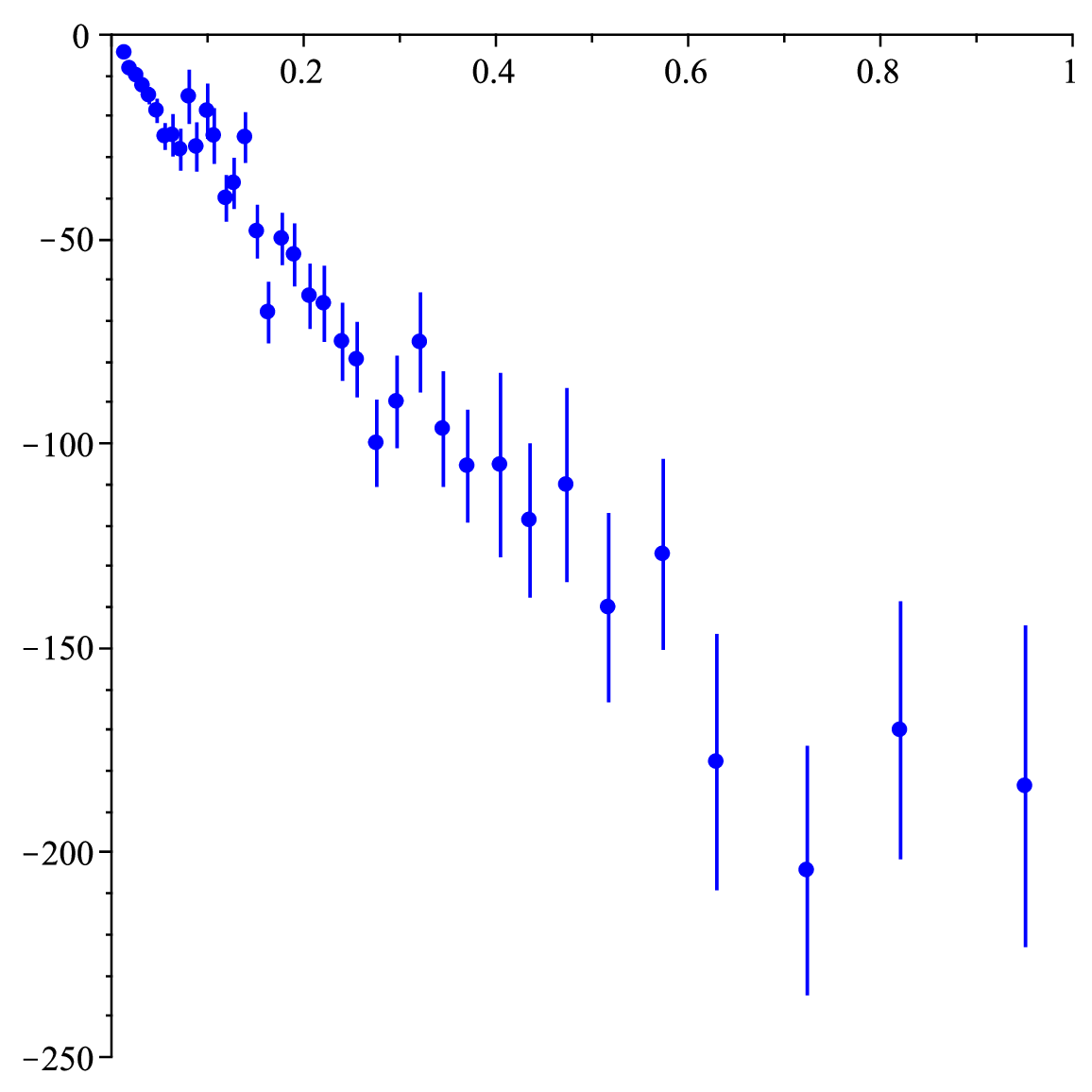}
\caption{
The distance-redshift relation from Pantheon type Ia supernovae
 (40 redshift bins) in blue dots with error bars.
The distance-redshift relation from BAO observations by BOSS and eBOSS
 in red dots with error bars in the inverse distance ladder.
The solid line is the prediction of the $\Lambda$CDM model with Planck CMB data
 with the error is inside the thickness of the line.
The left panel is the plot of whole available values of $z$,
 the middle panel is that enlarged with $z<1$,
 and the right panel is the Pantheon data subtracted by $\Lambda$CDM predictions.
}
\label{fig:LCDM-data}
\end{figure}

The observation of CMB by the Planck collaboration \cite{Planck:2018vyg} gives
 $\Omega_m = 0.315 \pm 0.007$ and $H_0 = 67.4 \pm 0.5 \,\, \rm{Km/s/Mpc}$,
 and we can obtain distance-redshift relation
 from eqs.(\ref{Distance}) and (\ref{HubbleParameter})
 with $\Omega_\Lambda = 1 - \Omega_m$ neglecting numerically small $\Omega_\gamma$.
In fig.\ref{fig:LCDM-data}
 the prediction of the $\Lambda$CDM model with Planck data
 is compared with the data from Pantheon supernovae and BAO.
This is the well-known result that
 it is consistent with the BAO with inverse distance ladder
 ($r_s^* = 147.09 \pm 0.26$ [Mpc] in \cite{Planck:2018vyg}),
 but strongly disfavored by Pantheon supernova data:
 the Hubble constant problem.
Especially in the region of $z<1$
 the prediction of the $\Lambda$CDM model with Planck CMB data
 gives systematically larger values of the distances of corresponding redshift values,
 which is corresponding to the difference beyond $4\sigma$ in $H_0$ values.
It should be also noted that
 the Pantheon supernova data are not totally consistent with BAO data
 in the inverse distance ladder.

\begin{figure}[t]
\centering
\includegraphics[width=50mm]{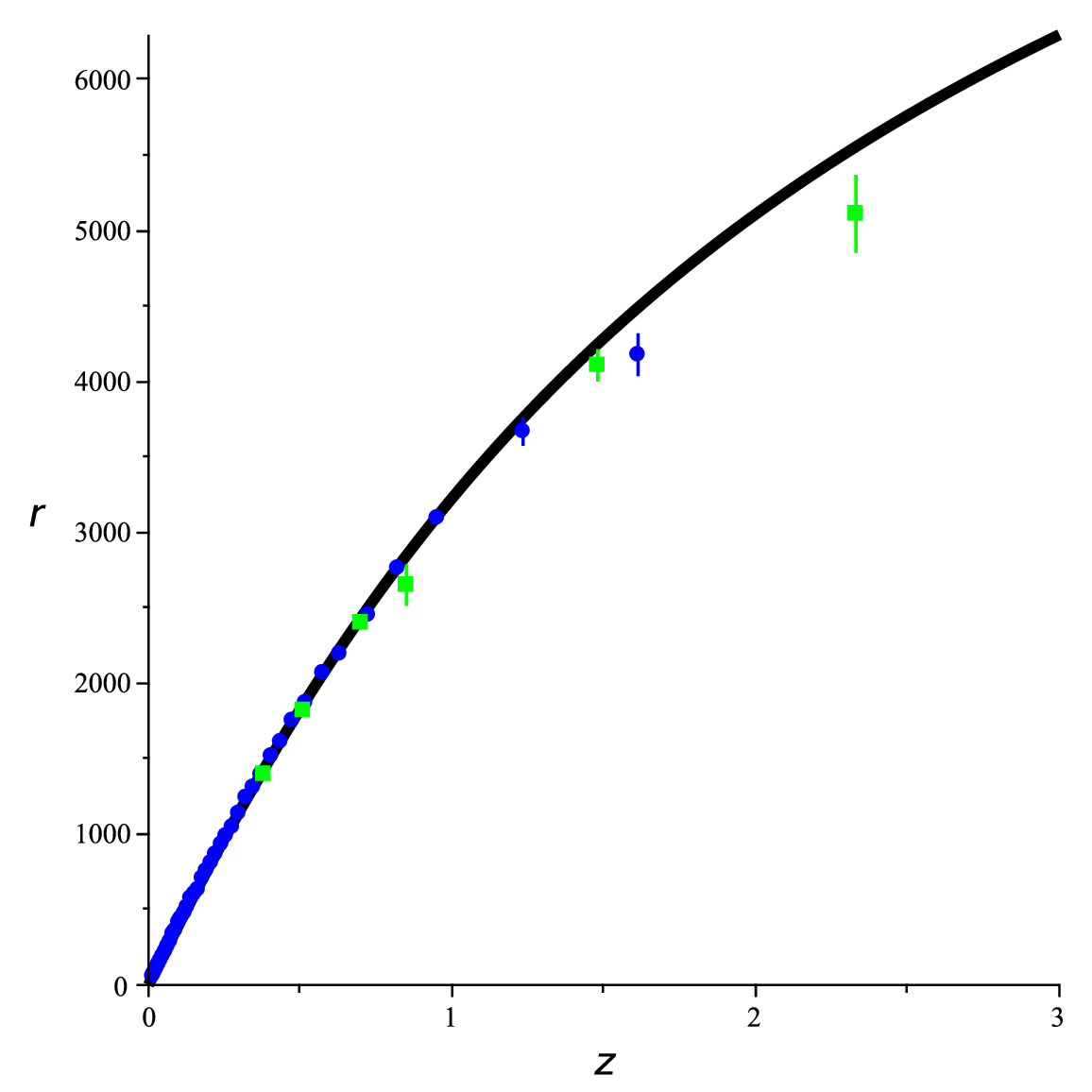}
\quad
\includegraphics[width=50mm]{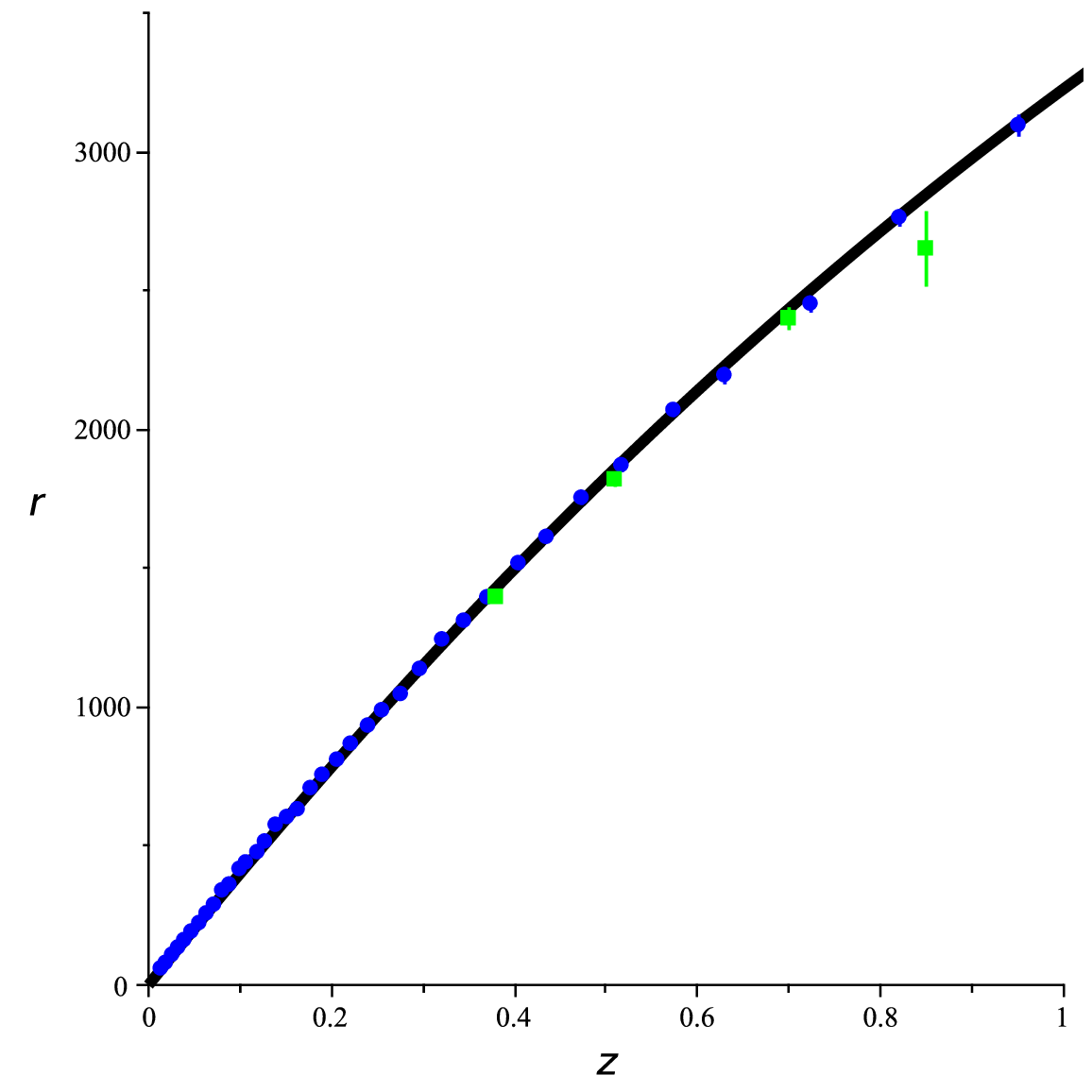}
\quad
\includegraphics[width=50mm]{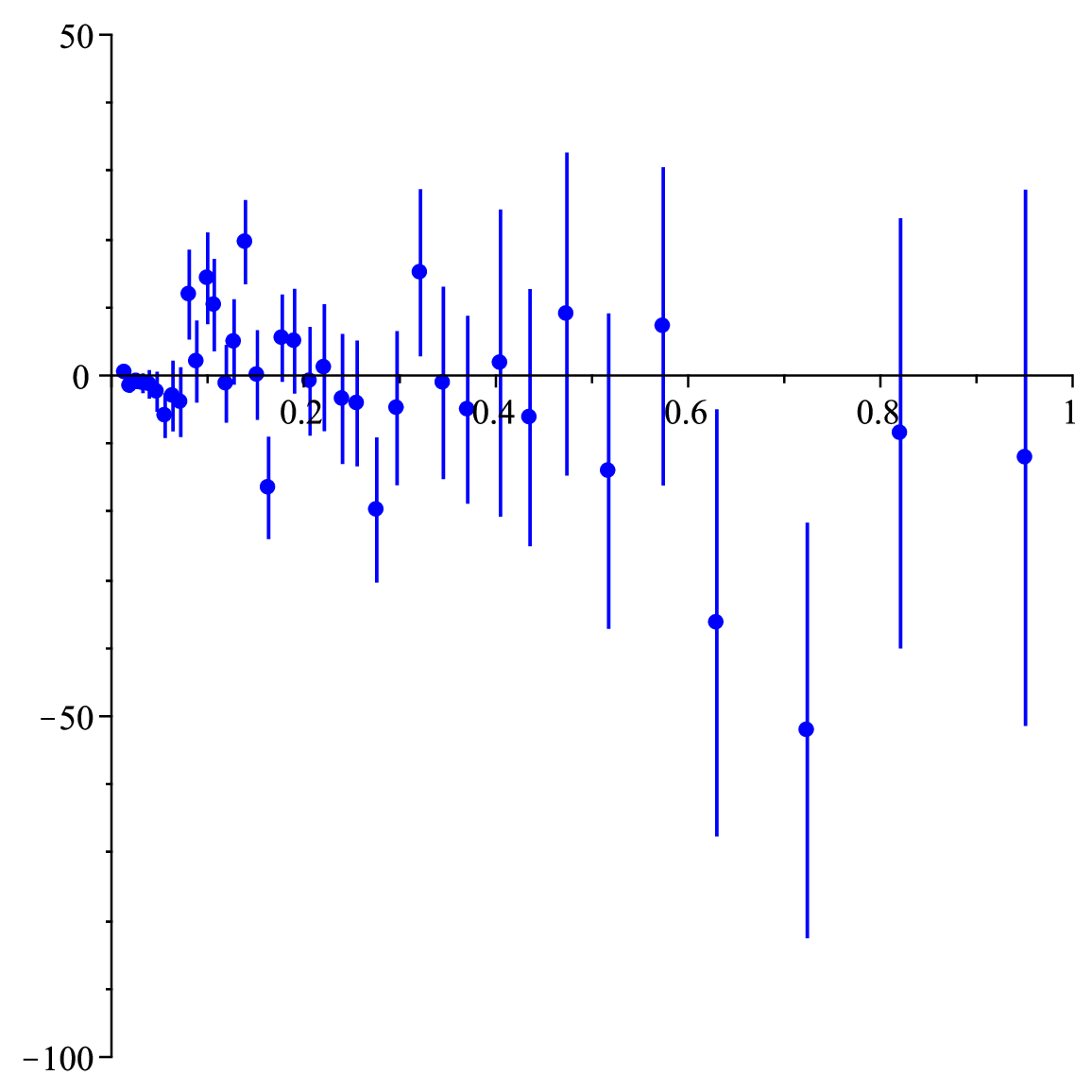}
\caption{
The distance-redshift relation from Pantheon type Ia supernovae
 (40 redshift bins) in blue dots with error bars.
The distance-redshift relation from BAO observations by BOSS and eBOSS
 in green dots with error bars in the local distance ladder.
The solid line is the prediction of the $\Lambda$CDM model with Planck CMB data
 of $\omega_b$ and $\omega_c$
 with artificially setting $H_0 = 73.2 \,\, \rm{Km/s/Mpc}$.
The error is inside the thickness of the line.
The left panel is the plot of whole available values of $z$,
 the middle panel is that enlarged with $z<1$,
 and the right panel is the Pantheon data subtracted by
 this artificial $\Lambda$CDM predictions.
}
\label{fig:LCDM-data-largeH0}
\end{figure}

If the value of $r_s^*$ was small
 by some new physics without changing the values of
 $\omega_b = 0.0204 \pm 0.0001$ and $\omega_c = 0.120 \pm 0.001$
 in Planck measurements so that $H_0 = 73.2 \,\, \rm{Km/s/Mpc}$,
 it would be highly consistent with the data
 as shown in fig.\ref{fig:LCDM-data-largeH0},
 where for BAO data
 we use $r_s^* = 135.9 \pm 3.2$ [Mpc]
 from the local distance ladder in \cite{eBOSS:2020yzd}.
Pantheon supernova data are totally consistent with BAO data in this case.
Though the data are reproduced rather well,
 an interesting observation is that
 there may be a problem in higher redshifts:
 the model tends to give too large value of $r(z)$ for larger $z$.
The results of future precise observations for higher redshifts by, for example,
 James Webb Space Telescope (JWST) for supernova observations
 and Euclid for BAO observations
 are expected to bring important information for the Hubble constant problem.
Since the behavior for larger redshift is determined by $\omega_m$
 we can also expect important information
 on the cold dark matter and structure formation.

Though this situation
 between the $\Lambda$CDM model with Pantheon supernova and BAO data,
 may be enough to suggest new physics for smaller $r_s^*$,
 we still need to investigate another possibility
 of non-trivial way of the expansion of the universe in $d_A^*$
 with time-dependent dark energy.

\section{Dark energy models and distance-redshift relation}
\label{sec:PhDE}

We assume that
 dark energy can be described as a perfect fluid
 which is defined by a equation of state $p_{\rm DE} = w_{\rm DE} \, \rho_{\rm DE}$,
 where $p_{\rm DE}$ and $\rho_{\rm DE}$ are pressure and energy density
 of the dark energy fluid.
A quick increase of the Hubble parameter in dark-energy dominated era ($z<0.3$)
 by some special equation of state of the dark energy
 (``hokey-stick equation of state'', for example)
 has already been excluded by Pantheon supernova data
 \cite{Benevento:2020fev,Camarena:2021jlr,Efstathiou:2021ocp}.
The dark energy models in this section are such that
 the time dependence of $\omega_\Lambda$,
 or vanishing dark energy at higher redshifts in eq.(\ref{HubbleParameter})
 increases the integrant of eq.(\ref{DistanceToLLS}),
 and the value of $h$ is increased to keep the value of $d_A^*$ unchanged.

\begin{figure}[t]
\centering
\includegraphics[width=50mm]{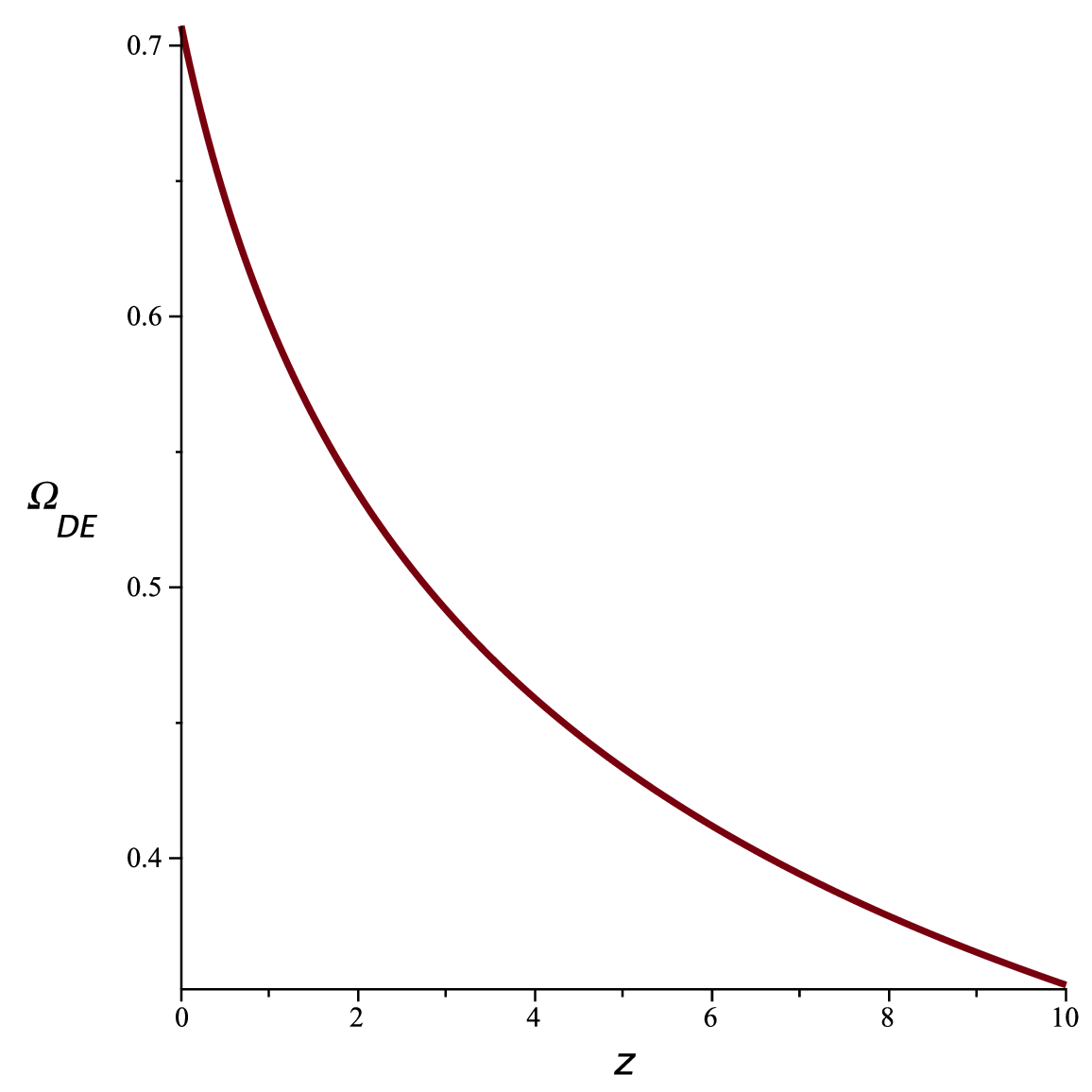}
\qquad
\includegraphics[width=50mm]{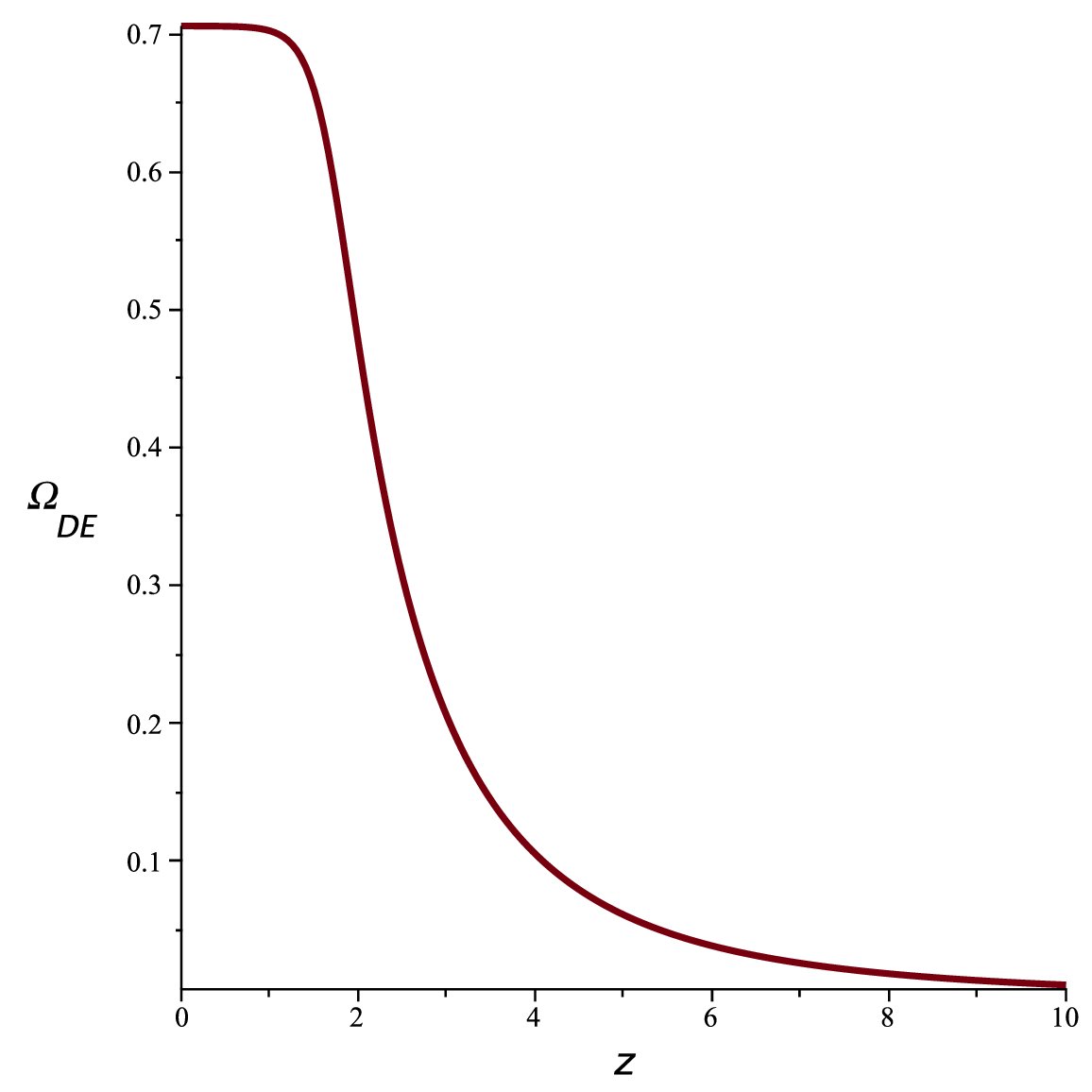}
\caption{
Left: the vacuum energy as a function of redshift
in ``generalized emergent dark energy model''.
Right: the vacuum energy as a function of redshift
in ``transitional dark energy model''.
}
\label{fig:DE-vs-z}
\end{figure}

First we consider the model,
 ``generalized emergent dark energy model'', proposed in \cite{Li:2020ybr}
 and examined with available data in \cite{Yang:2021eud}.
The dark energy equation of state is assumed to be
\begin{equation}
 w(z) = -1 - \frac{\Delta}{3\ln10}
 \left(
  1 + \tanh \left( \Delta \log_{10} \left(\frac{1+z}{1+z_t}\right) \right)
 \right),
\end{equation}
 and the resultant dark energy is
\begin{equation}
 \Omega_{\rm DE}(z) = \Omega_{{\rm DE}0} \, 
  \frac{1-\tanh(\Delta \log_{10}((1+z)/(1+z_t))}
       {1+\tanh(\Delta \log_{10}(1+z_t))}.
\end{equation}
Here, $\Delta$ is a parameter with $\Delta=0$ representing the $\Lambda$CDM model,
 $\Omega_{{\rm DE}0}=1-\Omega_m$,
 and $z_t$ is determined by solving $\Omega_{\rm DE}(z_t) = \Omega_m(1+z_t)^3$,
 which is the redshift value at `dark energy'-matter equality. 
The best fit values of parameters
 with data from Planck CMB, BAO, Pantheon and
 a measured value of $H_0 = 74.03 \pm 1.42 \,\, \rm{Km/s/Mpc}$ in \cite{Riess:2019cxk}
 are given in \cite{Yang:2021eud} as
\begin{equation}
 H_0 = 69.86^{+0.75}_{-0.74},
\qquad
 \Omega_m = 0.293^{+0.0065}_{-0.0067},
\qquad
 \Delta = 0.55^{+0.20}_{-0.21}.
\end{equation}
The non-zero best fit value of $\Delta$ indicates
 that the global fit of this model is better than that of the $\Lambda$CDM model.
Though the value of Hubble constant is larger
 than that of the $\Lambda$CDM model with CMB observations,
 it is much smaller than the typical value of direct measurements.
The dark energy as a function of redshift
 is given in the left panel in fig.\ref{fig:DE-vs-z}.
The dark energy has been emerged quickly and reaches a present value.
This model is purely phenomenological and no dynamics is specified behind it.

\begin{figure}[t]
\centering
\includegraphics[width=50mm]{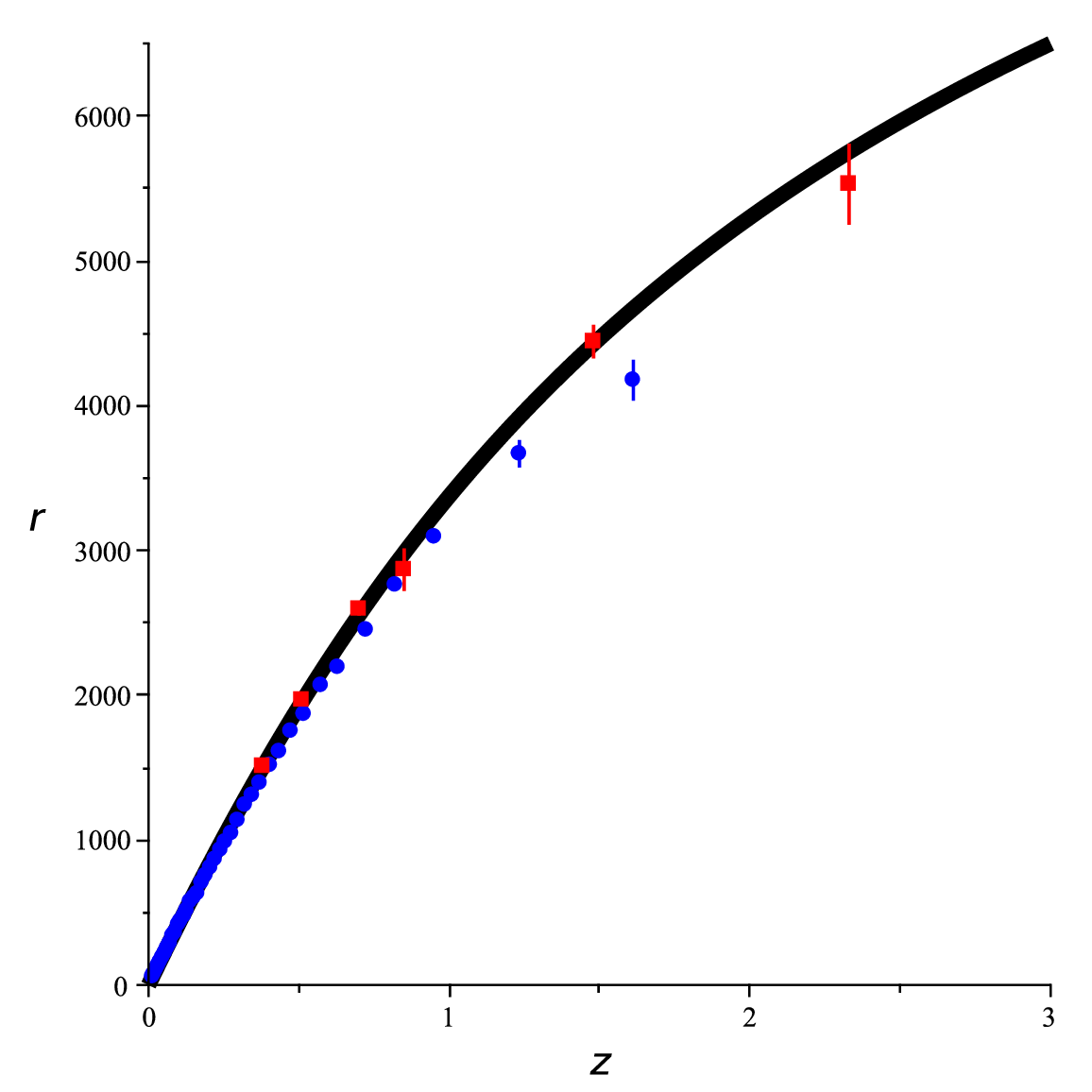}
\quad
\includegraphics[width=50mm]{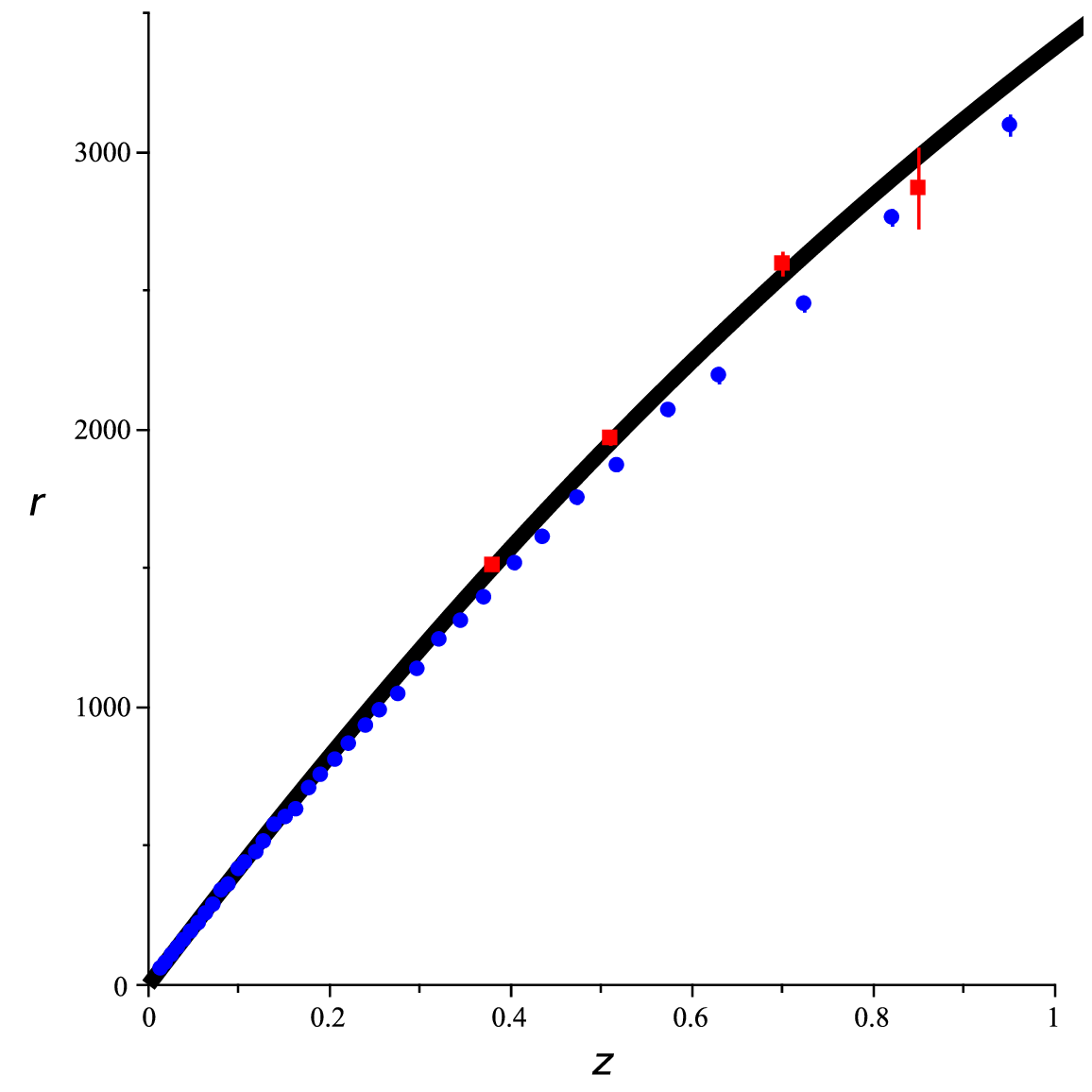}
\quad
\includegraphics[width=50mm]{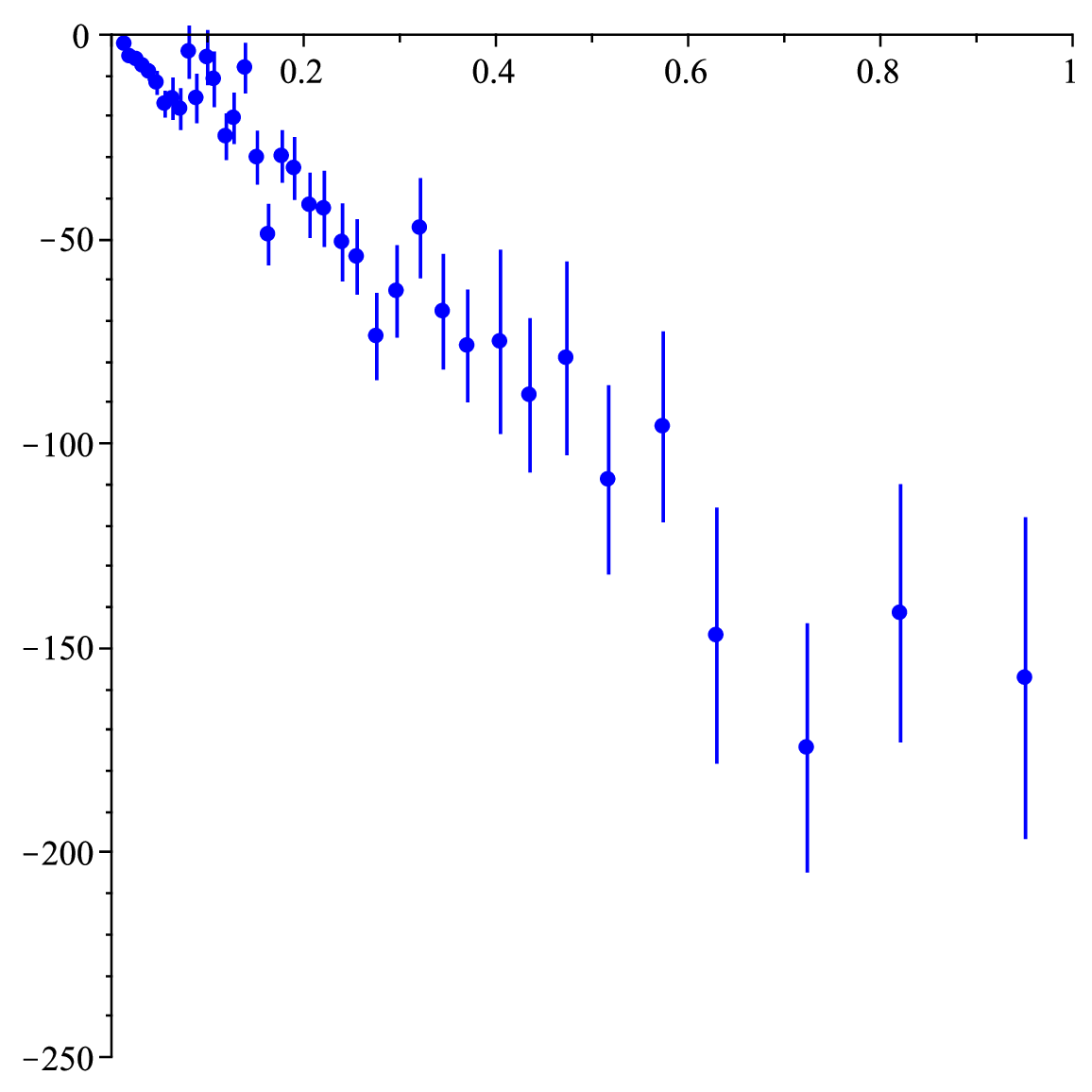}
\caption{
The distance-redshift relation from Pantheon type Ia supernovae
 (40 redshift bins) in blue dots with error bars.
The distance-redshift relation from BAO observations by BOSS and eBOSS
 in red dots with error bars in the inverse distance ladder.
The solid line is the prediction of
 ``generalized emergent dark energy model''
 with the error is inside the thickness of the line.
The left panel is the plot of whole available values of $z$
 the middle panel is that enlarged with $z<1$,
 and the right panel is the Pantheon data subtracted by the model predictions.
}
\label{fig:ED-data}
\end{figure}

In fig.\ref{fig:ED-data}
 the prediction of this ``generalized emergent dark energy model''
 is compared with the data from Pantheon supernovae and BAO.
The difference from fig.\ref{fig:LCDM-data} is that
 the fit is better for smaller values of redshift especially $z<0.5$.
The behavior at larger values of redshift is almost the same within errors.
Overall, however,
 this model is strongly disfavored by Pantheon supernova data
 at almost the same level for $\Lambda$CDM model with Planck data. 
In fact a Bayesian statistical analysis in \cite{Yang:2021eud} gives a result
 that the fit of this model only with the data from Planck and Pantheon
 is ruling out this model at more than $2\sigma$.
We need to be careful
 for the improvement of global fit which include all available data
 by changing model parameters,
 because the realization of a certain phenomenon may not be improved enough.

Next we consider the model,
 ``transitional dark energy model'', proposed in \cite{Zhou:2021xov}.
The dark energy equation of state is assumed to be
\begin{equation}
 w_{\rm DE}(z) = - 1 -\frac{1}{2} \left(\tanh\left(3(z-z_c)\right) + 1 \right),
\end{equation}
 where $z_c$ is the parameter which describes a transition period.
The best fit values of the parameters are given in \cite{Zhou:2021xov}
 with the data from Planck CMB observations,
 BOSS observations of BAO and redshift space distortions, Pantheon supernovae,
 and also with the values of Hubble constant in \cite{Riess:2020fzl}
 and of $S_8$ by Dark Energy Survey collaboration as
\begin{equation}
 H_0 = 69.16 \pm 0.76,
\qquad
 \omega_m = 0.1408 \pm 0.0010,
\qquad
 z_c = 1.667^{+0.31}_{-0.85}.
\end{equation}
In \cite{Zhou:2021xov}
 the goodness of fit of this model
 is better than that of $\Lambda$CDM with Planck data.
Again,
 though the value of Hubble constant
 is larger than that of the $\Lambda$CDM model with CMB observations,
 it is much smaller than the typical values of direct measurements.
The redshift dependence of dark energy
 is numerically calculated as the right panel of fig.\ref{fig:DE-vs-z}.
The dark energy is almost constant at small redshifts $z < z_c$,
 and it quickly dumps for larger redshift values.
This behavior could be ideal
 to realize present accelerated expansion of the universe
 with maximally reduced integrant of eq.(\ref{DistanceToLLS})
 for large value of $h$.

\begin{figure}[t]
\centering
\includegraphics[width=50mm]{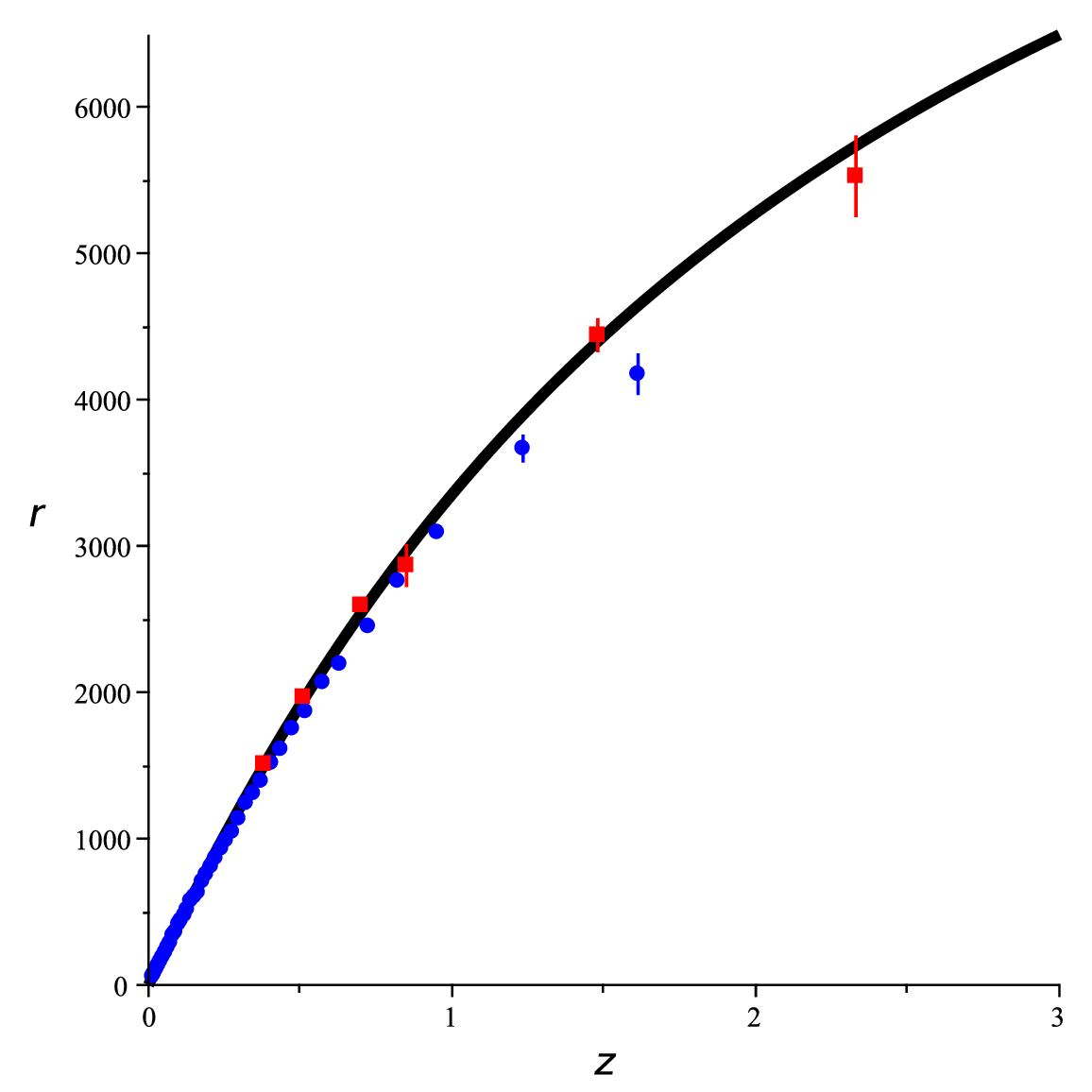}
\quad
\includegraphics[width=50mm]{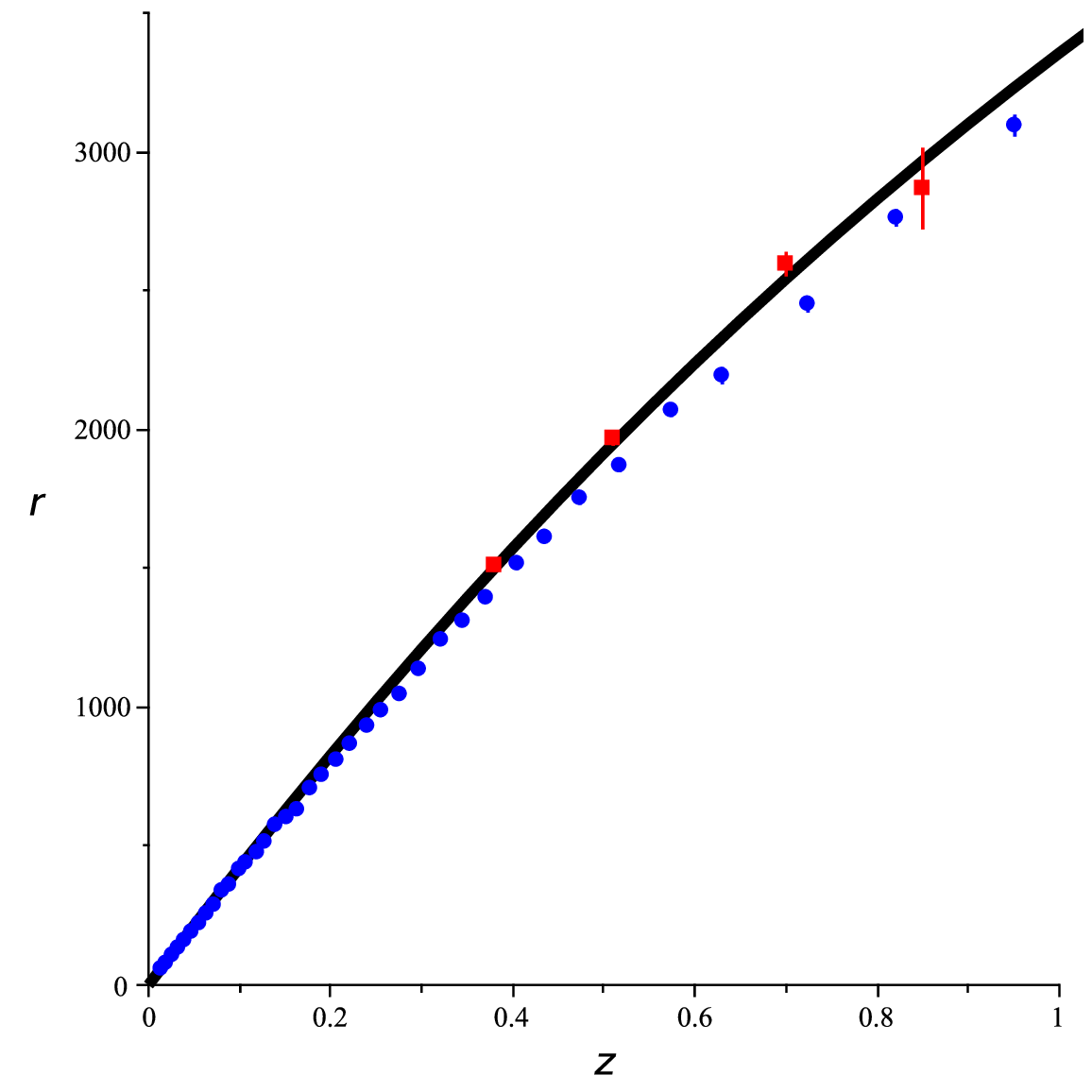}
\quad
\includegraphics[width=50mm]{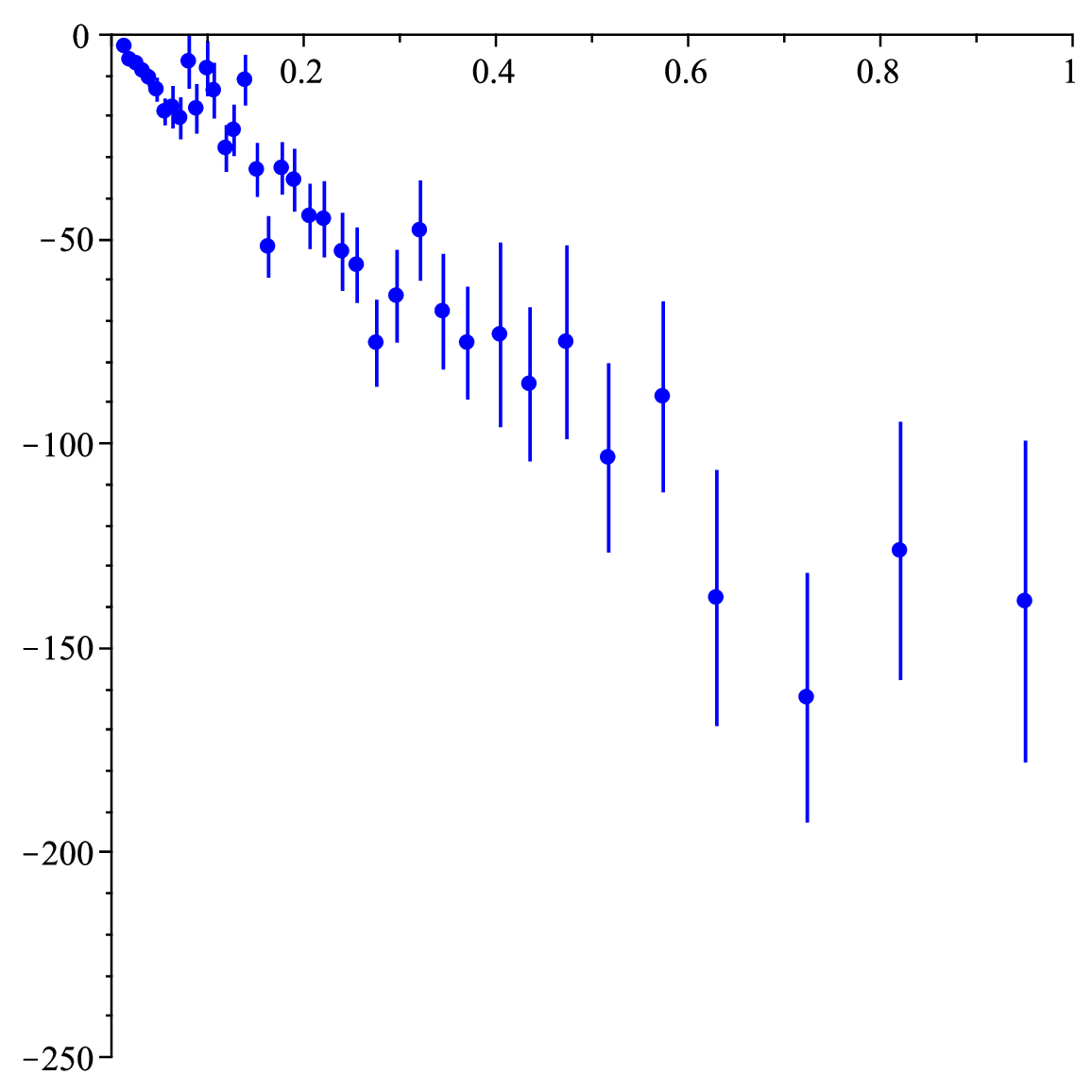}
\caption{
The distance-redshift relation from Pantheon type Ia supernovae
 (40 redshift bins) in blue dots with error bars.
The distance-redshift relation from BAO observations by BOSS and eBOSS
 in red dots with error bars in the inverse distance ladder.
The solid line is the prediction of
 ``transitional dark energy model''
 with the error is inside the thickness of the line.
The left panel is the plot of whole available values of $z$
 the middle panel is that enlarged with $z<1$,
 and the right panel is the Pantheon data subtracted by the model predictions.
}
\label{fig:TD-data}
\end{figure}

In fig.\ref{fig:TD-data}
 the prediction of this ``transitional dark energy model''
 is compared with the data from Pantheon supernovae and BAO.
The difference from fig.\ref{fig:LCDM-data} is that
 the prediction of the model becomes closer to the Pantheon supernova data
 in all the redshift region.
However, again,
 this model is strongly disfavored by Pantheon supernova data
 at almost the same level for the $\Lambda$CDM model with Planck data.

Since the dark energy
 should realize the present accelerated expansion of the universe,
 the present value of the energy density, $\Omega_\Lambda=1-\Omega_m$,
 is strongly constrained
 assuming flat space-time neglecting small contribution of radiation.
The modification of dark energy
 can give large direct effect only in a short redshift period $z < 0.3$,
 and the effect almost disappears for larger redshift in matter dominated era.
The only effect for larger redshifts 
 is the difference of the values of $\omega_m$
 from the global fit with available data including the modification of dark energy.
Therefore, as two models in this section show,
 we can not reproduce distance-redshift relation from Pantheon supernova data
 by introducing redshift-dependent dark energy.
The strategy of introducing non-trivial way of the expansion of the universe
 in $d_A^*$ with time-dependent dark energy is not successful.
It is apparent with present data that
 changing the shape of the function of $r(z)$ is not important,
 but changing the overall normalization of $r(z)$ is required.\footnote{
 Allowing the transition to negative values of vacuum energy at higher redshifts
  is beyond the scope of our arguments
  \cite{Akarsu:2019hmw,Akarsu:2021fol,Akarsu:2022typ}.}

\section{Discussions and conclusions}
\label{sec:conclusions}

There are two strategies to solve the Hubble constant problem:
 to realize smaller value of $r_s^*$ or
 to make $d_A^*$ giving larger values of $h$
 by changing the expansion of the universe after recombination.
The present late-time data
 of Pantheon supernovae and BAO by BOSS and eBOSS strongly suggest
 the first strategy, namely,
 the value of sound horizon at recombination
 should be reduced than that predicted by the $\Lambda$CDM model
 whose parameters are fixed by CMB observations.
The second strategy
 with time-dependent dark energy
 has been strongly disfavored by Pantheon supernova data.
Actually
 the plots in fig.\ref{fig:LCDM-data-largeH0} strongly suggest
 the dark energy as the cosmological constant,
 which has been supported also by the observations of
 weak lensing and galaxy clustering (see \cite{DES:2022ccp} for a recent result).
We conclude that
 there must be some new physics beyond the $\Lambda$CDM model
 which naturally gives a smaller value of $r_s^*$.
The arguments in previous sections
 may seem naive without mentioning any statistical significances, 
 but they give clear information
 which could be overlooked in the global fit with all the available data.
We can not accept a model,
 if it is strongly disfavored by a part of data,
 even if the other data can be reasonably reproduced.

This conclusion
 may be changed by future observations for higher redshifts $z > 1$
 by JWST and Euclid, for example.
Even the plots in fig.\ref{fig:LCDM-data-largeH0}
 potentially have a problem of fit for larger redshifts,
 where the universe is matter dominated and
 the value of physical energy density of matter $\omega_m$ is important.
Depending on the results by future observations
 we may expect to be able to obtain some non-trivial information on
 cold dark natter.
It is also very important
 to check the possible systematic uncertainty in the Pantheon supernova observation
 by future observations (see \cite{Yuan:2022edy}, for example).

It is well-known that
 it is not an easy task to reduce the value of $r_s^*$
 by introducing some early-time new physics
 beyond the $\Lambda$CDM model (see \cite{Vagnozzi:2023nrq}, for example).
The observed perturbation of CMB is beautifully reproduced
 by a simple $\Lambda$CDM model with a combination of many effects.
The almost scale-invariant primordial scalar perturbations
 are developed into baryon acoustic oscillations in baryon-photon plasma
 under the effect of dark matter and free-streaming neutrinos.
At the time of recombination
 the photon temperature perturbations arise from many sources:
 the original temperature perturbations of baryon-photon plasma,
 the perturbations through the Doppler effect
 by velocity perturbations of baryon-photon plasma,
 the perturbations of gravitational potential
 producing photon temperature perturbations by Sachs-Wolfe effect,
 and Silk dumping reducing shorter wavelength of the photon temperature perturbations.
It is also important that
 the time-dependent gravitational potential
 in matter dominant era before recombination
 produces further photon temperature perturbations by integrated Sachs-Wolfe effect.
Any early-time new physics, which reduces the value of $r_s^*$,
 affects some of them and changes the feature of CMB power spectrum.
For example in \cite{Vagnozzi:2021gjh}
 it is investigated in detail
 how the results of the integrated Sachs-Wolfe effect
 is affected by the early-time new physics.
In early dark energy models
 (see \cite{Poulin:2018cxd,Niedermann:2020dwg} for typical proposal
  and [2, 3] for more complete lists),
 which are typical early-time new physics for smaller values of $r_s^*$,
 the integrated Sachs-Wolfe effect becomes week,
 and we need to increase the energy density of cold dark matter $\omega_c$,
 which increase siginificantly the late-time matter perturbation $S_8$
 beyond the constraints by present observations.
Note that
 the recent result of observation \cite{Kilo-DegreeSurvey:2023gfr}
 gives the value of $S_8$ which is consistent
 with that of the $\Lambda$CDM model with Planck data.

This problem may be beyond that
 can be solved by simply adding something to the $\Lambda$CDM model.
It may require
 to reconsider simultaneously the physics of inflation \cite{Cruz:2022oqk},
 or even
 to change the fundamental law of the particle physics \cite{Sekiguchi:2020teg}.

\section*{Acknowledgments}

This work was supported in part by JSPS KAKENHI Grant Number 19K03851.

\end{document}